\documentclass{acmart}

\renewcommand\footnotetextcopyrightpermission[1]{} % removes footnote with conference information in first column
\usepackage{hyperref}
\usepackage{graphicx}
\usepackage{booktabs} 
\usepackage{lineno}
\usepackage{amsmath}
\usepackage{calligra}
\usepackage{tablefootnote}
\usepackage{fontawesome}
\usepackage{supertabular}
\usepackage{adjustbox}
\usepackage{comment}
\usepackage[xcolor]{changebar}
\usepackage{makecell}
\usepackage{multirow}
\usepackage{subcaption}
\usepackage{svg}
\usepackage{graphics}

\usepackage[xcolor]{changebar}

\usepackage{lipsum}% http://ctan.org/pkg/lipsum

\usepackage{enumitem}
\usepackage[T1]{fontenc}

\usepackage{tikz}
\usepackage{url}
\usepackage{mathtools, cuted}
\usepackage{booktabs} % For formal tables
\usepackage{svg}
\usepackage{comment}
\usepackage{xcolor}
\usepackage{multirow}
\usepackage{multicol}
\usepackage{enumitem}
\usepackage{fontawesome}
\usepackage{adjustbox}
\usepackage{graphicx,mathtools}
\usepackage{supertabular}
\usepackage[labelfont=bf]{caption}
\usepackage{amsmath}
\usepackage{amsthm}
\usepackage{bm}
\usepackage[misc,geometry]{ifsym}

\usepackage{breqn}

\newcommand{\sizecirle}{0.8ex}
\newcommand{\rc}{\tikz\fill[red] (0,0) circle (\sizecirle);}
\newcommand{\bc}{\tikz\fill[blue] (0,0) circle (\sizecirle);}

\DeclareRobustCommand{\llp}{\text{\bc}}
\DeclareRobustCommand{\lcp}{\text{\rc}}

\begin{document}

% Main title of the paper
% Main title of the paper
\title{Case Study: The Impact of Location on Bias in Search Results}

\author[1]{Gizem Gezici}
\affiliation{%
  \institution{Huawei Turkey R\&D Center}
  \country{ Istanbul, Turkey}
}

%\vspace{-4em}
\begin{abstract}

In this work, we aim to investigate the impact of location (different countries) on bias in search results. For this, we use the search results of Google and Bing in the UK and US locations. The query set is composed of controversial queries obtained from ProCon.org that have specific ideological leanings as conservative or liberal. Please note that this study is a follow-up work of~\cite{gezici2021evaluation}. In the previous work, researchers analyse search results in terms of stance and ideological bias with rank and relevance based measures. Yet, in the scope of this work, by using the query subset of~\cite{gezici2021evaluation} we examine the effect of location on the existence of bias as well as the magnitude of bias difference between Bing and Google. Note that this study follows a similar evaluation procedure. Our preliminary results show that location might affect the retrieval performance of search engines as well as the bias in the search results returned by Bing and Google towards the controversial queries.
%Engineering, and Mathematics) and NON-STEM fields of education. Gender is a research area by itself in social sciences which is beyond the scope of this work. In this respect, for annotating the perceived gender of the narrator of an instructional video we used only a crude classification of gender into Male, and Female. Then, for analysing perceived gender bias we utilised bias measures that have been inspired by search platforms and further incorporated rank information into our analysis. Our preliminary results demonstrate that there is a significant bias towards the male gender on the returned YouTube educational videos, and the degree of bias varies when we compare STEM and NON-STEM queries. Finally, there is a strong evidence that rank information might affect the results.

\end{abstract}

\maketitle

\section{Research Questions}
\label{sec:intro}
In this work, the impact of location on web search bias is investigated as well. For this, online search bias is analysed for the UK and US versions of Bing and Google. Specifically, this study aims to answer the following research questions.

The first research question is:

\begin{description}
    \item[RQ1:] On a conservative-liberal ideology space, do search engines return~\emph{biased} SERPs and if so; are these biases~\emph{significantly different} from each other towards controversial topics?
\end{description}

In order to answer RQ1, the degree of deviation of the ranked SERPs from an~\emph{equal} representation of different ideologies, is measured. Further, the magnitude of ideological bias in Google and Bing is compared.

The second research question is:

\begin{description}
    \item[RQ2:] On a conservative-liberal ideology space, do different geolocations affect the existence of~\emph{ideological} bias in search engines?
\end{description}

Specifically in the scope of this chapter, the effect of location is examined. Initially, the existence of bias is analysed in the UK version of Bing (Google) and the US version of Bing (Google) to check if different locations affect the existence of ideological bias in each search engine. 

The last research question is:

\begin{description}
    \item[RQ3:] On a conservative-liberal ideology space, do different geolocations affect the magnitude of~\emph{ideological} bias difference in search engines?
\end{description}

In addition to investigating the effect of location on the existence of bias, it is examined whether different locations influence the magnitude of bias. For this, the bias of the UK version of Bing (Google) and the US version of Bing (Google) is compared in terms of level of bias they show.

\section{Experimental Setup}
In this section, the steps of the evaluation procedure to examine the location effect on web search bias are described.

\subsection{Dataset}
\label{sec:search_bias_loc_dataset}
For the dataset crawling, the subset of the controversial queries obtained from ProCon were used. In the scope of this work, only the queries of the controversial topics that has ideological leanings were used for the analysis of location. In this work, to investigate the impact of location, only the queries that could be used in ideological bias were selected for the bias analysis. For this purpose, all the 38 topics with their queries (topic questions) were leveraged to crawl the top-10 SERPs of Bing and Google using the locations of the UK and US.

For crawling the SERPs, the news channels of UK and US Google and Bing search engines were used. To avoid any personalisation effect, news search results were collected in~\emph{incognito mode}. Thus, the retrieved SERPs are not tailored to any particular user, but (presumably) to all UK/US users. Each topic question, query, was submitted to UK/US Google and Bing's news search engines using UK/US proxies. Note that the news channel do not show any sponsored results in the SERPs.
First, the URLs of the retrieved SERPs were crawled for the same topic question (query) in order to minimise time lags between search engines in the same location and specifically for this chapter in different locations as well. Thus, news SERPs of Bing and Google were automatically crawled using the UK and US proxies in parallel. In this way, the time gap between Bing-US and Google-US and their UK counterparts was attempted to be minimised. Since it has been known that news channel is less dynamic than the default web channel, the minimised time lags would probably little effect on the search results. Following that, the crawled URLs were used to extract the textual contents of the top-10 documents.

\subsection{Crowd-sourcing Campaigns}
\label{sec:search_bias_loc_crowdsourcing}
To label the stance of each document with respect to the topic questions (queries), crowd-sourcing was used and MTurk was chosen as a platform. To ensure high-quality crowd-labeling, the following task properties were specified in this platform. Although the majority of the issues being relevant to the US, in the scope of this work for the annotation of UK SERPs, crowd workers from the UK and for the US SERPs crowd workers from the US were hired for the annotation. Additionally, skilled and experienced professionals were attempted to be recruited by establishing the following criteria: The approval rate for Human Intelligence Tasks (HITs), i.e. single, self-contained task for a worker, should be better than 95\%. Per each HIT, the wage was set as 0.02\$, i.e. in the scope of web search bias each HIT contained 5 annotations instead of 1 annotation, and a time limit was 30 minutes. Three crowd-workers judged each document. Although a similar crowd sourcing procedure was attempted to apply, lower inter-rater agreement scores were obtained as $0.3215$ and $0.2979$ for the UK and US SERPs annotations respectively. This is probably because, a detailed iterative process could not be fulfilled due to time and budget constraints.

\subsection{Quantifying Bias}
\label{sec:search_bias_loc_quantifyingbias}
The bias of four search engines, the UK and US versions of Bing and Google, is quantified by analysing their news channels.
While the one-sample t-test is applied to check the existence of bias in search engines separately, the two-tailed paired t-test is used to check whether the magnitude of bias difference between two search engines is statistically significant. In the scope of location analysis, Bonferroni correction has also been applied since there are many hypotheses to be checked using t-tests. Bonferroni correction is used for multiple hypothesis testing and in the context of this bias analysis, there are 36 hypotheses in total. Hence, without the Bonferroni correction, with the significance level, $ \alpha = .05$ and 36 hypotheses, the probability of identifying at least one significant result due to chance is around $0.84$. Note that for the significance level where $\alpha = .05$, and with the Bonferroni correction new $\alpha = .00138$ and Bonferroni correction rejects the null hypothesis for each p-value ($p_i$) if $p_i$ <= $.00138$ instead of $.05$. For the significance level where $\alpha = .01$, new $\alpha = .00028$ and for $\alpha = .001$, new $\alpha = .000028$ and so on.

\subsection{Results}
\label{sec:search_bias_loc_results}

The impact of location has been investigated mainly in two ways, first on overall ideological bias results, then on ideological bias of each search engine separately.

%\subsubsection{The Impact of Location on Overall Ideological Bias Results}
%\label{sec:ideologicalbias_location}
%
Prior to examining the existence of bias, first Google and Bing's retrieval performances were measured for the UK and US locations independently.

In Table~\ref{tab:performance_location_UK} and Table~\ref{tab:performance_location_US} it is observed that Bing and Google show similar retrieval performances -- the two-tailed paired t-tests computed on retrieval scores are statistically not significant for the UK and US locations. This is verified across all three IR evaluation measures. Nonetheless, it is observed that the US versions of Bing and Google show higher retrieval performances than the UK.
For this, in Table~\ref{tab:performance_location_engine1} and Table~\ref{tab:performance_location_engine2}, the retrieval performances of engine 1 and engine 2 were assessed for the UK and US locations. The results show that the US versions of both search engines show higher retrieval performances than their UK counterparts. The two-tailed paired t-tests computed on the retrieval scores are statistically significant for engine 1 and engine 2 in Table~\ref{tab:performance_location_engine1} and Table~\ref{tab:performance_location_engine2} respectively. This is verified across all three IR evaluation measures.

Following that, it is determined whether Google and Bing return biased results in the UK and US locations. In Table~\ref{tab:ideological_UK}, both UK search engines are ideologically biased towards conservative (all MB scores are positive) -- one sample t-tests computed on MB and MAB scores are statistically significant. The bias scores for the ideological leanings of conservative and liberal bias scores cancelled each other out, thus MBs show lower scores than MABs. Similarly, both US search engines seems to be ideologically biased in Table~\ref{tab:ideological_US}, however one sample t-tests computed on MB scores are statistically not significant but computed on MAB scores are statistically significant. This is probably because, the bias scores for the ideological leanings of conservative and liberal bias scores cancelled each other out. Unlike the UK search engines, neither of the US engines are biased. In addition, there is no difference in the magnitude of bias -- two-tailed paired t-tests computed on MBs and MABs are statistically not significant for both the UK and US search engines. This is verified across all three IR evaluation measures.

Apart from these, it has also been investigated if the same search engine (engine 1 or engine 2) shows similar level of bias in different locations. In Table~\ref{tab:ideological_engine1}, the UK version of engine 1 seems to be more biased towards conservative than its US counterpart. Yet, two-tailed paired t-tests computed on MB scores are statistically not significant with Bonferroni correction for the measures $P@10$, $RBP$, and $DCG@10$. In terms of MABs, both the UK and US versions of engine 1 show similar level of absolute bias -- two-tailed paired t-tests computed on MAB scores are statistically not significant and this is confirmed by all three IR evaluation measures. For engine 2, in Table~\ref{tab:ideological_engine2}, two-tailed paired t-tests computed on MB scores are statistically not significant for $P@10$, while statistically significant for $RBP$, and $DCG@10$ due to Bonferroni correction ($p-values$ = $.0113$, $.0013$, and $ .0006$ for $P@10$, $RBP$, and $DCG@10$ respectively). Similar to engine 1, both the UK and US versions of engine 2 also show similar level of absolute bias -- two-tailed paired t-tests computed on MAB scores are statistically not significant and this is confirmed by all three IR evaluation measures.

\begin{figure}[!t]
\centering
%\hspace{-1.8em}
\begin{minipage}{.45\textwidth}
  \centering
    %\hspace{-1em}
    {\includegraphics[width=1\textwidth,trim={2.5cm 6.5cm 4.5cm 6.5cm},clip]{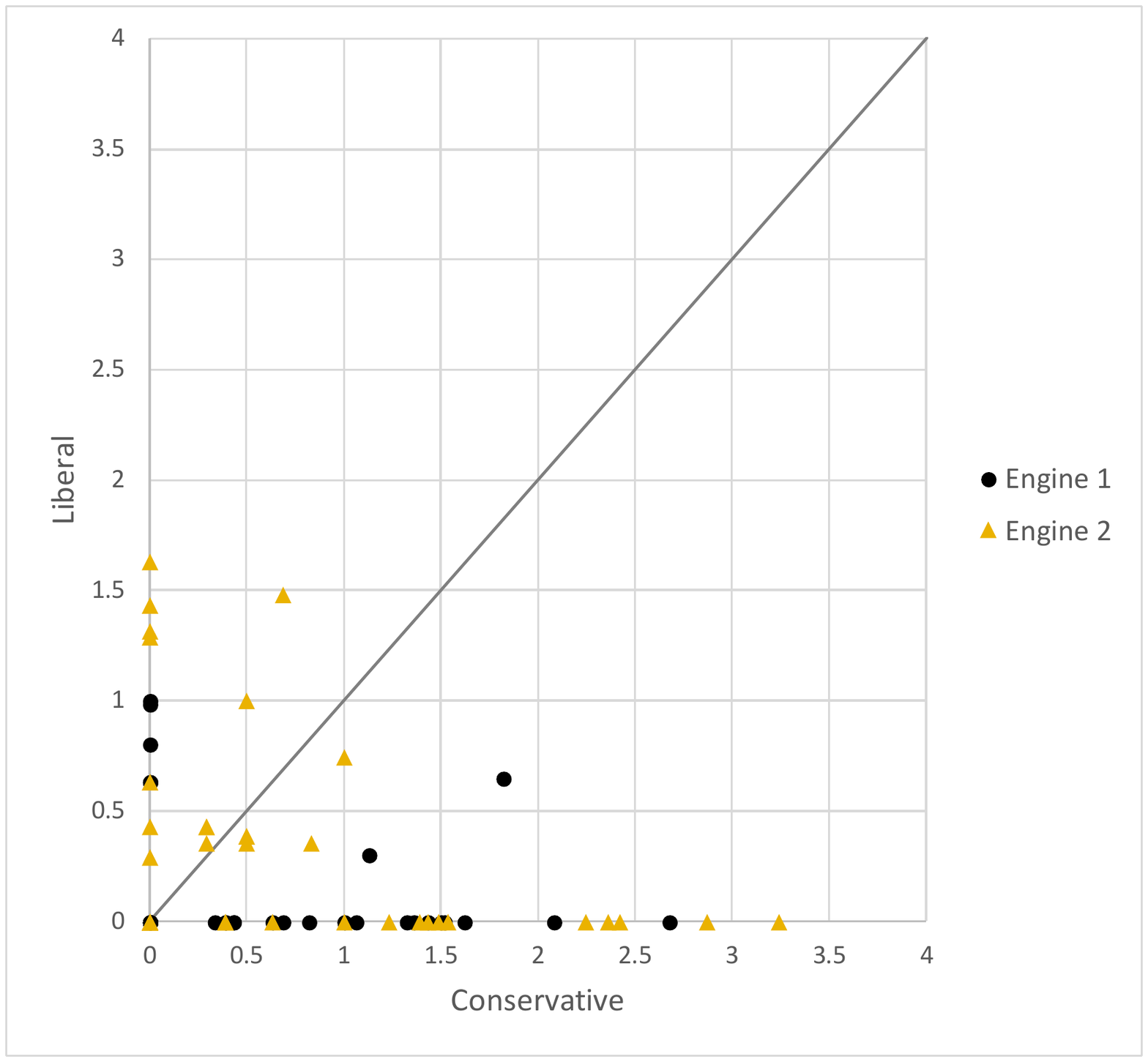}}
    \caption{$DCG_{\lcp}@10$ against $DCG_{\llp}@10$ measured on ideological leanings of UK -- black points for engine 1 and yellow points for engine 2}
    \label{fig:dcg_ideologyUK}
\end{minipage}
\hspace{2.8em}
\begin{minipage}{.45\textwidth}
    \centering
    \vspace{0.1em}
    %\hspace{-1em}
    {\includegraphics[width=1\textwidth,trim={2.5cm 6.5cm 4.5cm 6.5cm},clip]{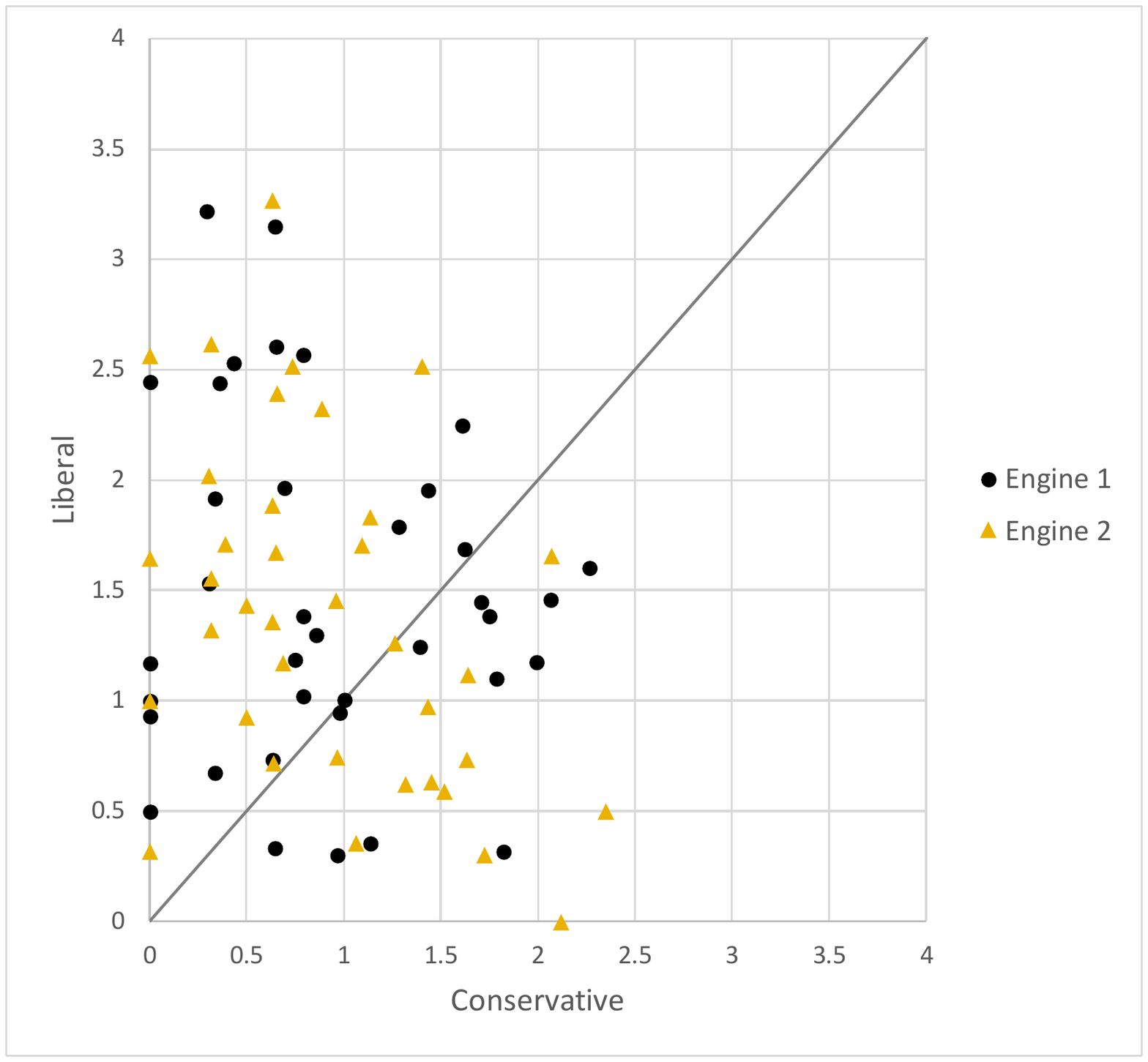}}
    \caption{$DCG_{\lcp}@10$ against $DCG_{\llp}@10$ measured on ideological leanings of US -- black points for engine 1 and yellow points for engine 2}
    \label{fig:dcg_ideologyUS}
\end{minipage}
%\vspace{-1.5em}
\end{figure}

\begin{figure}[!t]
\centering
%\hspace{-1.8em}
\begin{minipage}{.45\textwidth}
  \centering
    %\hspace{-1em}
    {\includegraphics[width=1\textwidth,trim={2.5cm 5.25cm 2cm 5cm},clip]{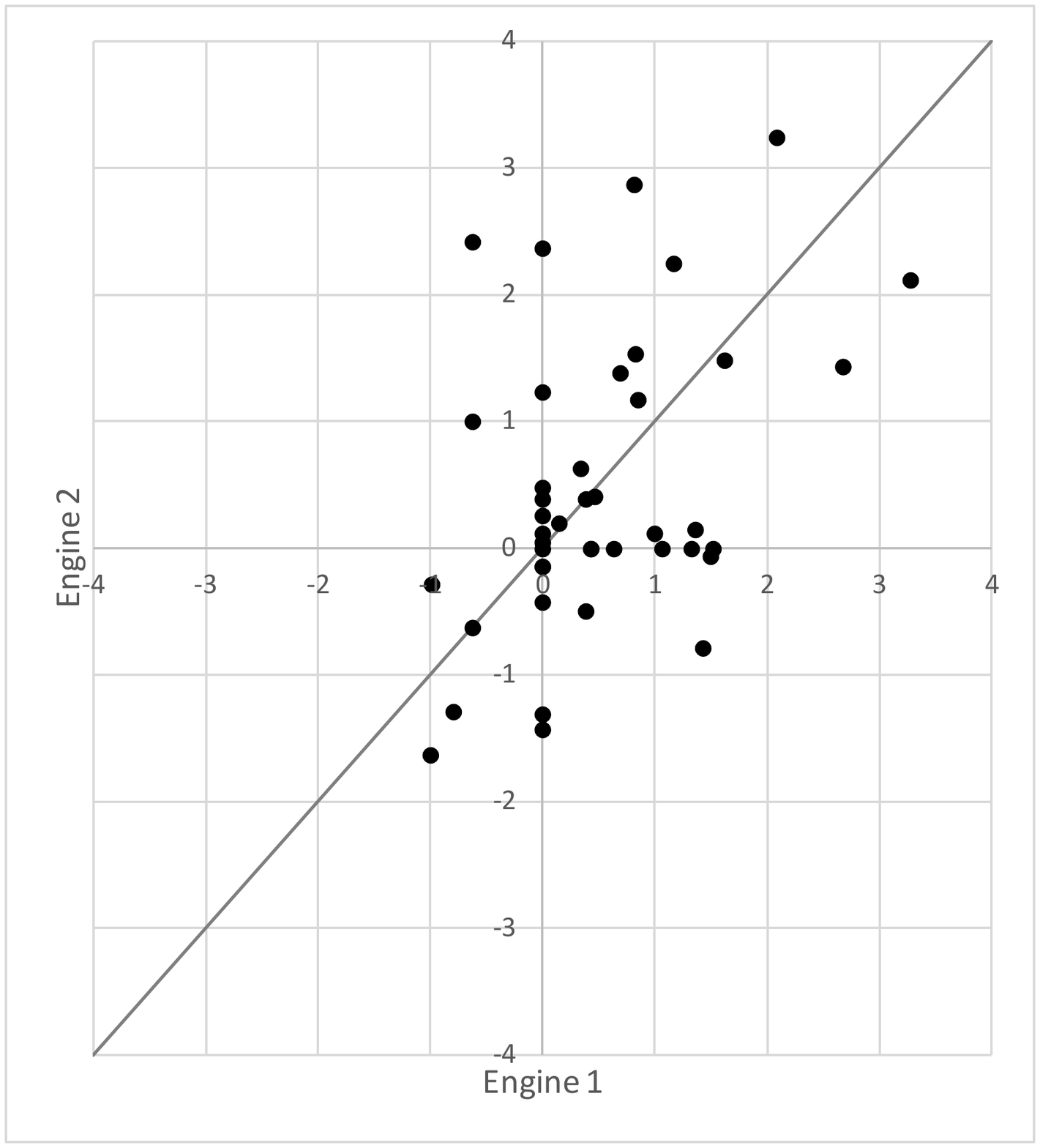}}
    \caption{$\beta_{DCG@10}$ measured on UK leanings, where positive is $\lcp$ and negative is $\llp$}
    \label{fig:dcgdiff_ideologyUK}
\end{minipage}
\hspace{2.8em}
\begin{minipage}{.45\textwidth}
    \centering
    \vspace{0.1em}
    %\hspace{-1em}
    {\includegraphics[width=1\textwidth,trim={2.5cm 5.25cm 2cm 5cm},clip]{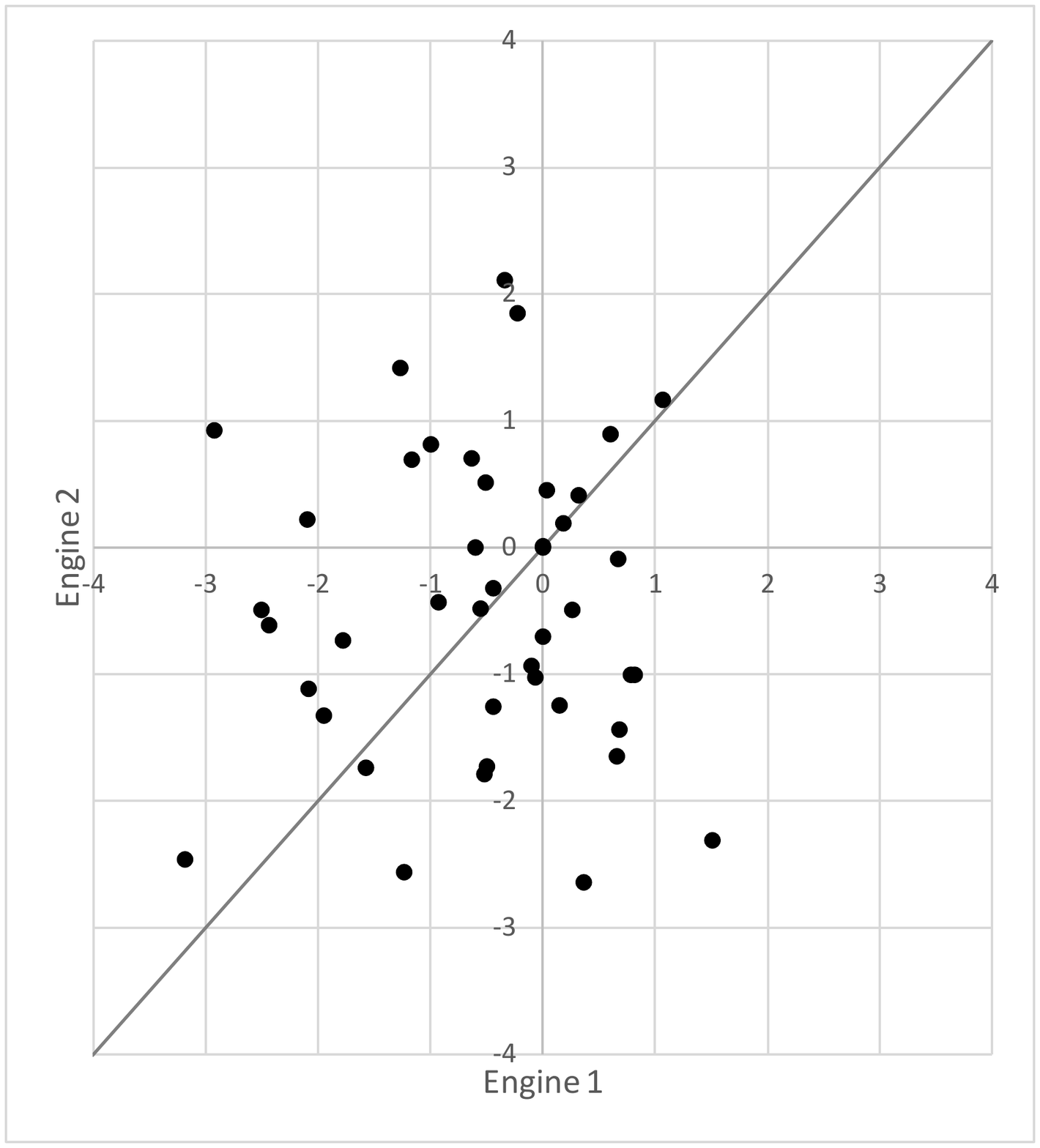}}
    \caption{$\beta_{DCG@10}$ measured on US leanings, where positive is $\lcp$ and negative is $\llp$}
    \label{fig:dcgdiff_ideologyUS}
\end{minipage}
%\vspace{-1.5em}
\end{figure}

In Figure~\ref{fig:dcg_ideologyUK}, the distribution of query-specific SERPs over the conservative-liberal ideological spectrum for the DCG@10 measure in the UK is depicted. The x-axis represents the conservative-ideological score ($DCG_{\lcp}@10$), whereas the y-axis represents the liberal-ideological score ($DCG_{\llp}@10$). Each point represents a query's overall SERP score. The black points represent the SERPs that engine 1 retrieved, whereas the yellow points represent the SERPs that engine 2 retrieved. Similarly, in Figure~\ref{fig:dcg_ideologyUS}, the distribution of topic-specific SERPs over the conservative-liberal ideological spectrum in the US is displayed.
The overall ideological bias score ($\beta_{DCG@10}$) of SERPs for each query evaluated on the two UK search engines is visualized in Figure~\ref{fig:dcgdiff_ideologyUK}. The x-axis represents engine 1, while the y-axis represents engine 2. Positive coordinates denote topics whose SERPs are skewed towards the conservative leaning, whereas negative coordinates denote topics whose SERPs are skewed towards the liberal leaning. Similar to Figure~\ref{fig:dcgdiff_ideologyUK}, Figure~\ref{fig:dcg_ideologyUS} displays the overall ideological bias score ($\beta_{DCG@10}$) of SERPs for each query evaluated on the US search engines.

%search engines' performances

\begin{table}[!t]
    %\vspace{1em}
    \centering
    % Engine 1 is better (Bing)
    \caption{The UK search engines' performance, as determined by the p-values of a two-tailed paired t-test performed on engines 1 and 2.}
    \begin{tabular}{cccc}
        \hline\hline
        & P@10 & RBP & DCG@10 \\
        \hline
        Engine 1 & 0.6027 & 0.5896 & 2.9203 \\
        Engine 2 & 0.6649 & 0.6170 & 3.0993\\
        \hline
        p-value & $ > 0.05$ & $ > 0.05$ & $ > 0.05 $ \\
        \hline\hline
    \end{tabular}
    \label{tab:performance_location_UK}
    %\vspace{1em}
\end{table}

\begin{table}[!t]
    %\vspace{1em}
    \centering
    % Engine 1 is better (Bing)
    \caption{The US search engines' performance, as determined by the p-values of a two-tailed paired t-test performed on engines 1 and 2.}
    \begin{tabular}{cccc}
        \hline\hline
        & P@10 & RBP & DCG@10 \\
        \hline
        Engine 1 & 0.9730 & 0.8734 & 4.4305 \\
        Engine 2 & 0.9838 & 0.8790 & 4.4691 \\
        \hline
        p-value & $ > 0.05$ & $ > 0.05$ & $ > 0.05 $ \\
        \hline\hline
    \end{tabular}
    \label{tab:performance_location_US}
    %\vspace{1em}
\end{table}

%%%%%%%ENGINE1 & 2 LOC-WISE COMPARISON

\begin{table}[!t]
    %\vspace{1em}
    \centering
    % Engine 1 is better (Bing)
    \caption{The location-wise performance of engine 1, as determined by the p-values of a two-tailed paired t-test performed on the UK engine 1 and  US engine 1.}
    \begin{tabular}{cccc}
        \hline\hline
        & P@10 & RBP & DCG@10 \\
        \hline
        Engine 1 (UK) & \textbf{0.6027} & \textbf{0.5896} & \textbf{2.9203} \\
        Engine 1 (US) & \textbf{0.9730} & \textbf{0.8734} & \textbf{4.4305} \\
        \hline
        p-value & $ \textbf{ < 0.0001} $ & $ \textbf{ < 0.0001} $ & $ \textbf{ < 0.0001} $ \\
        \hline\hline
    \end{tabular}
    \label{tab:performance_location_engine1}
    %\vspace{1em}
\end{table}

\begin{table}[!t]
    %\vspace{1em}
    \centering
    % Engine 1 is better (Bing)
    \caption{The location-wise performance of engine 2, as determined by the p-values of a two-tailed paired t-test performed on the UK engine 2 and  US engine 2.}
    \begin{tabular}{cccc}
        \hline\hline
        & P@10 & RBP & DCG@10 \\
        \hline
        Engine 2 (UK) & \textbf{0.6649} & \textbf{0.6170} & \textbf{3.099} \\
        Engine 2 (US) & \textbf{0.9838} & \textbf{0.8790} & \textbf{4.4691} \\
        \hline
        p-value & $ \textbf{ < 0.0001} $ & $ \textbf{ < 0.0001} $ & $ \textbf{ < 0.0001} $ \\
        \hline\hline
    \end{tabular}
    \label{tab:performance_location_engine2}
    %\vspace{1em}
\end{table}

%%%%Ideological Bias

\begin{table*}[!t]
    \centering
    \caption{The UK search engines' ideological bias, as evaluated by the p-values of a two-tailed paired t-test performed on engine 1 and engine 2.
    }
    \hspace{-0.7em}\begin{tabular}{cccccccc}
        \hline\hline
        & & P@10 & & RBP & & DCG@10 & \\
        \hline
        \multirow{3}{*}{MB} 
        & Engine 1 & 0.1108 & & 0.1214 & & 0.5740 & \\	
        & Engine 2 & 0.1027 & & 0.1339 & & 0.6260 &\\
        \cline{2-8}
        & p-value & $>0.05$ & & $>0.05$ & & $>0.05$ &\\
        \hline
        \multirow{3}{*}{MAB} 
        & Engine 1 & 0.1378 && 0.1573 && 0.7205 & \\
        & Engine 2 & 0.1622	&& 0.1873 && 0.8829 &  \\
        \cline{2-8}
        & p-value & $>0.05$ & & $>0.05$ & & $>0.05$ &\\
        \hline\hline
    \end{tabular}
    \label{tab:ideological_UK}
    %\vspace{1.2em}
\end{table*}

\begin{table*}[!t]
    \centering
    \caption{The US search engines' ideological bias, as evaluated by the p-values of a two-tailed paired t-test performed on engine 1 and engine 2.
    }
    \hspace{-0.7em}\begin{tabular}{cccccccc}
        \hline\hline
        & & P@10 & & RBP & & DCG@10 & \\
        \hline
        \multirow{3}{*}{MB} 
        & Engine 1 & -0.0027 & & 0.0107 & & -0.0065 & \\	
        & Engine 2 & -0.0405 & & -0.0693 & & -0.2718 &\\
        \cline{2-8}
        & p-value & $>0.05$ & & $>0.05$ & & $>0.05$ &\\
        \hline
        \multirow{3}{*}{MAB} 
        & Engine 1 & 0.1811 && 0.2039 && 0.9247 & \\
        & Engine 2 & 0.1865 && 0.2354 && 1.0623 &  \\
        \cline{2-8}
        & p-value & $>0.05$ & & $>0.05$ & & $>0.05$ &\\
        \hline\hline
    \end{tabular}
    \label{tab:ideological_US}
    %\vspace{1.2em}
\end{table*}

\begin{table*}[!t]
    \centering
    \caption{The location-wise ideological bias of engine 1, as evaluated by the p-values of a two-tailed paired t-test performed on the UK engine 1 and US engine 1.
    }
    \hspace{-0.7em}\begin{tabular}{cccccccc}
        \hline\hline
        & & P@10 & & RBP & & DCG@10 & \\
        \hline
        \multirow{3}{*}{MB} 
        & Engine 1 (UK) & 0.1108 & & 0.1214 & & 0.5740 & \\	
        & Engine 1 (US) & -0.0027 & & 0.0107 & & -0.0065 &\\
        \cline{2-8}
        &  p-value & $ > 0.05 $ & & $ > 0.05 $ & & $ > 0.05 $ &\\
        \hline
        \multirow{3}{*}{MAB} 
        & Engine 1 (UK) & 0.1378 && 0.1573 && 0.7205 & \\
        & Engine 1 (US) & 0.1811 && 0.2039 && 0.9247 &  \\
        \cline{2-8}
        & p-value & $>0.05$ & & $>0.05$ & & $>0.05$ &\\
        \hline\hline
    \end{tabular}
    \label{tab:ideological_engine1}
    %\vspace{1.2em}
\end{table*}

\begin{table*}[!t]
    \centering
    \caption{The location-wise ideological bias of engine 2, as evaluated by the p-values of a two-tailed paired t-test performed on the UK engine 2 and US engine 2.
    }
    \hspace{-0.7em}\begin{tabular}{cccccccc}
        \hline\hline
        & & P@10 & & RBP & & DCG@10 & \\
        \hline
        \multirow{3}{*}{MB} 
        & Engine 2 (UK) & 0.1027 & & \textbf{0.1339} & & \textbf{0.6260} & \\	
        & Engine 2 (US) & -0.0405 & & \textbf{-0.0693} & & \textbf{-0.2718} &\\
        \cline{2-8}
        & p-value & $ \textbf{< 0.05} $ & & $ \textbf{< 0.05} $ & & $ \textbf{< 0.05}$ &\\
        \hline
        \multirow{3}{*}{MAB} 
        & Engine 2 (UK) & 0.1622 && 0.1873 && 0.8829 & \\
        & Engine 2 (US) & 0.1865 && 0.2345 && 1.0623 &  \\
        \cline{2-8}
        & p-value & $>0.05$ & & $>0.05$ & & $>0.05$ &\\
        \hline\hline
    \end{tabular}
    \label{tab:ideological_engine2}
    %\vspace{1.2em}
\end{table*}

%%%%%%%%%%%%%%%%

%%%FIGURES

\subsection{Concluding Discussion}
Before evaluating the possibility of bias in SERPs, the retrieval performances of two distinct search engines are compared. As seen in Table~\ref{tab:performance_location_UK} and Table~\ref{tab:performance_location_US} both search engines work effectively in the UK and US respectively, but engine 2 outperforms engine 1 yet the difference is statistically not significant. This is verified by the three IR evaluation measures. Also, it is observed that the US versions of both search engines work better than their UK counterparts.
For this, the retrieval of the same search engine (Google or Bing) is compared in the UK and US locations to check whether the location affects the retrieval performance, or not. In Table~\ref{tab:performance_location_engine1}, in terms of retrieval performance the US version of engine 1 outperforms its UK counterpart -- the two tailed paired t-test provides statistically significant results and it is verified across the three IR evaluation measures. Likewise, in Table~\ref{tab:performance_location_engine2} the US version of engine 2 outperforms its UK counterpart as well.

%left-lower-right-upper

%
Then, it is determined whether search engines return biased results in terms of ideology leanings (\textbf{RQ1}) and, if this is the case, it is then determined if the search engines exhibit the same level of bias (\textbf{RQ1}), suggesting the difference between engines is statistically not significant.
In Table~\ref{tab:ideological_UK}, all MB scores are positive, and the UK search engines appear to be biased in favour of the conservative leaning with regard to~\textbf{RQ1}.
The one-sample t-test is used to determine the presence of stance bias, that is, whether the true mean is different from zero, as discussed in Section~\ref{sec:search_bias_loc_quantifyingbias}. The one-sample t-test computed on MB scores is statistically significant for both search engines. Nonetheless, in order to answer the~\textbf{RQ1}, two-tailed paired t-test computed on MBs are statistically significant which means that the UK versions of the search engines show similar level of bias.
On the basis of MAB scores, it is clear that both engines exhibit an absolute bias. The two-tailed t-test demonstrates that the difference between the two engines is statistically not significant. This means that both search engines show similar level of absolute bias.
Unlike, in Table~\ref{tab:ideological_US}, all MB scores are negative meaning that the US versions of the search engines seem to be biased towards the liberal learning with respect to~\textbf{RQ1}. Yet, the one-sample t-test computed on MB scores is statistically not significant so neither of the US search engines are biased. The one-sample t-test computed on MAB scores, it is observed that both engines exhibit an absolute bias. The two-tailed paired t-test computed on MAB scores is statistically not significant, thus both search engines exhibit similar level of bias~(\textbf{RQ1}). Regarding the~\textbf{RQ2}, the UK versions of both search engines are biased towards conservative, while neither of the US search engines are biased.

Regarding the~\textbf{RQ3}, in Table~\ref{tab:ideological_engine1}, the two-tailed paired t-test computed on MBs and MABs is statistically not significant which means that location does not affect the magnitude of bias that the engine 1 exhibits. Unlike, in Table~\ref{tab:ideological_engine2}, the two-tailed paired t-test computed on MBs are statistically significant for $RBP$, and $DCG@10$ meaning that location affects the magnitude of bias in the case of the engine 2. Based on MAB scores, the engine 2 exhibits the same level of bias irrespective of the location.
In Figure~\ref{fig:dcg_ideologyUK}, both engine 1 and engine 2 seems to be biased towards the conservative leaning -- the query points are generally appear to be far away from the trendline. Unlike, in Figure~\ref{fig:dcg_ideologyUS}, the query points are more dispersed and most of them are close to the trendline -- neither of the search engines seem to be clearly biased. In Figure~\ref{fig:dcgdiff_ideologyUK}, both search engines appear to be biased towards conservative, whereas in Figure~\ref{fig:dcgdiff_ideologyUS}, the query points are more dispersed and there is no visible bias towards a specific ideological leaning. These interpretations are consistent with our aforementioned conclusions inferred from the results.

%\newpage
\bibliographystyle{plain}
\bibliography{main}
%\newpage

\end{document}